\documentclass[aps,reprint,amsmath,amssymb]{revtex4-2}

\usepackage{graphicx}
\usepackage{times}
\usepackage[colorlinks=true,
            linkcolor=blue,
            citecolor=blue,
            urlcolor=blue]{hyperref}

\begin{document}

\title{Nonlinear visco-elasto-plastic model for the stress-strain and stress-relaxation in metallic glass}

\author{Ankit Singh}
\affiliation{Department of Physics ``A. Pontremoli", University of Milan, Via Celoria 16, 20133 Milan, Italy}
\email{ankit.singh@unimi.it; alessio.zaccone@unimi.it}

\author{Oleg Gendelman}
\affiliation{Faculty of Mechanical Engineering,
Technion – Israel Institute of Technology, Haifa, Israel}
\email{ovgend@technion.ac.il}

\author{Alessio Zaccone}
\affiliation{Department of Physics ``A. Pontremoli", University of Milan, Via Celoria 16, 20133 Milan, Italy}
\email{alessio.zaccone@unimi.it}

\author{Valeriy V. Ginzburg}
\affiliation{Department of Chemical Engineering and Materials Science, Michigan State University, East Lansing, Michigan 48824, USA}
\email{ginzbur7@msu.edu}

\date{\today}

\begin{abstract}
We develop a nonlinear visco-elasto-plastic (VEP) model for the stress-strain response of metallic glasses by connecting the deformation dynamics with the material relaxation time obtained from independent stress-relaxation measurements. The material is modeled as a Maxwell-type element, with a nonlinear elastic spring and a stretched-exponential dashpot whose behavior is determined from the separate stress-relaxation data. The model captures the effects of temperature and strain rate on the initial elastic response, stress overshoot, post-yield softening, and steady flow. We independently determine the reference material-time function $X_0(\varepsilon)$ from the stress-relaxation response and the nonlinear material-time variable $X(\varepsilon)$ from the elastic-viscoplastic strain decomposition based on the stress-strain response.
 We introduce an Eyring-type function or material-time ratio $\phi(\varepsilon)=X(\varepsilon)/X_0(\varepsilon)$ which quantifies the change of relaxation dynamics during nonlinear deformation. Our model provides a simple connection between structural relaxation and yielding in metallic glasses.
\end{abstract}

\maketitle

\section{Introduction}
Mechanical response of liquids and amorphous materials can be understood from the classical picture introduced by Frenkel, where the response is controlled by the competition between elastic deformation and structural relaxation \cite{frenkel8kinetic,trachenko2013duality}. At times shorter than the structural relaxation time $\tau$, the material behaves more like an elastic solid, whereas at longer times structural rearrangements become possible and the response becomes viscous \cite{Zaccone2023}. In the Maxwell-Frenkel picture, the relaxation time is related to the viscosity and the high-frequency shear modulus through $\tau=\eta/G_{\infty}$~\cite{trachenko2013duality,RevModPhys_dyre_2006}, where $\eta$ is the viscosity and $G_{\infty}$ is the infinite-frequency shear modulus. Thus, the relaxation time gives the natural time scale that separates solid-like and liquid-like mechanical behaviors \cite{safarik2005shear,yu2015unified}.

In supercooled liquids and metallic glasses, the disordered atomic structure gives rise to a complex mechanical response that strongly depends on temperature and deformation rate \cite{schuh2007mechanical,hufnagel2016deformation}. Deformation generally starts with an elastic regime, followed by nonaffine strain-softening, yielding, stress overshoot, eventually reaching a Newtonian-like steady plastic or viscous-flow regime \cite{lu2003deformation,anand2007constitutive,PhysRevB.90.140203}. These features depend strongly on temperature, strain rate, and the structural state of the glass. One of the most important signatures of the nonlinear response is the stress overshoot observed during constant-strain-rate deformation \cite{lu2003deformation}. The stress first increases, reaches a maximum, and then decreases toward a steady-flow value \cite{lu2003deformation}. The magnitude of the overshoot changes strongly with temperature and strain rate, indicating a competition between the imposed deformation time scale and the intrinsic structural-relaxation time of the material \cite{lu2003deformation,PhysRevLett_1999}.

Several theoretical approaches have been proposed to describe the deformation of metallic glasses, including free-volume models \cite{spaepen1977microscopic}, shear-transformation-zone descriptions \cite{argon1979plastic,PhysRevE_1998,PhysRevE_STZ_2008,jiang_2015origin,lin2023mean}, energy-barrier model \cite{jiang_2015origin,mutneja_2026microscopic}, nonaffine deformation models \cite{PhysRevB.90.140203,wang_2022PRL,dong_2023non,Zaccone_2011,laurati_2017long,PhysRevMaterials_2023} and phenomenological elastic-viscoplastic constitutive models \cite{anand2005theory,anand2007constitutive,henann2008constitutive,wang2016maxwell}. These approaches are able to describe many features of yielding and flow, such as stress overshoot and strain softening \cite{PhysRevB.90.140203,anand2007constitutive,jiang_2015origin}. However, the direct connection between the relaxation dynamics measured independently in the time domain and the nonlinear stress-strain response is still not usually introduced explicitly in the constitutive description.

At the same time, the relaxation dynamics of metallic glasses is generally non-Debye \cite{qiao2012kinetics, qiao2013relaxation}. Stress-relaxation and mechanical-spectroscopy measurements are often described by stretched-exponential relaxation, characterized by a relaxation time $\tau$ and a stretching exponent $\beta$ \cite{qiao2013relaxation,wang2014evolution}. The strong temperature dependence of $\tau$ provides a natural time scale that can be compared directly with the time scale of the applied deformation \cite{PhysRevLett_1999,PhysRevLett_2017,PhysRevLett_2019}. In this way, the relaxation process can play a central role in determining how much of the imposed deformation is stored elastically and how much is relaxed through viscous motion.

A useful way to connect relaxation with nonlinear deformation is through the concept of material time \cite{narayanaswamy1971model}. In this description, the evolution of the material is measured by an internal material clock rather than directly by laboratory time \cite{dyre2015narayanaswamy}. This idea has been widely used in the description of structural relaxation in glassy systems and provides a simple way to treat both non-exponential and nonlinear relaxation processes \cite{narayanaswamy1971model,dyre2015narayanaswamy}.

In this work, building on the Frenkel-Kontorova approximation of the underlying energy landscape \cite{chaikin,FrenkelKontorova1938}, we develop a nonlinear visco-elasto-plastic description of the stress-strain response of metallic glasses by directly connecting the deformation dynamics with the material relaxation time obtained from independent stress-relaxation measurements. The total strain is decomposed into elastic and viscoplastic contributions within a Maxwell-type picture (the latter can also be derived from first-principles based on microscopic nonaffine motions as shown in Ref. \cite{PhysRevE.108.044101}). The nonlinear elastic response describes the stress overshoot and yielding, while the viscoplastic contribution is formulated through a material-time description constrained by the measured relaxation dynamics. We utilized a similar principle recently to investigate the brittle-to-ductile transition (BDT) in polymers \cite{soft_matter_2026}. Here, we apply the new comprehensive model to metallic-glass data at different temperatures and strain rates and show that it captures the initial elastic response, stress overshoot, post-yield softening, and steady-flow regime  \cite{lu2003deformation}. We independently determine the reference material-time function $X_0(\varepsilon)$ from the stress-relaxation response and the nonlinear material-time variable $X(\varepsilon)$ from the elastic-viscoplastic strain decomposition based on the stress-strain response.  We also calculate the Eyring-type function \cite{eyring1936viscosity}$ ~\phi(\varepsilon)=X(\varepsilon)/X_0(\varepsilon)$, which quantifies how the material clock changes as the system enters the nonlinear yield regime.

\section{Model and Theoretical Description}

\subsection{Maxwell-type decomposition}

We consider the uniaxial deformation of a glassy material in terms of an
elastic element and a viscous element connected in series. The two elements
therefore carry the same stress, while the total strain is decomposed into
elastic and viscoplastic contributions \cite{Bird1987},
\begin{equation}
\sigma_{\rm el}=\sigma_{\rm vp}=\sigma,
\qquad
\varepsilon
=
\varepsilon_{\rm el}
+
\varepsilon_{\rm vp}.
\label{eq:maxwell}
\end{equation}
For the elastic element, the linear response at small deformation is
$\sigma_{\rm el}=E\varepsilon_{\rm el}$, where $E$ is the Young's modulus.
For a Newtonian dashpot,
$\sigma_{\rm vp}=\eta\,d\varepsilon_{\rm vp}/dt$, with
$\eta=E\tau$, where $\tau$ denotes the characteristic relaxation time.
For a constant-strain-rate experiment,
$\varepsilon(t)=\dot{\varepsilon}t$.

\subsection{Nonlinear elastic response}

To describe yielding, we assume that the material initially evolves inside a
local minimum of the potential-energy landscape \cite{FrenkelKontorova1938,PhysRevLett.95.195501,tang2020bulk}. The latter is typically approximate as a periodic function to reflect the ``granularity" of matter at the atomic scale. The elastic stress is then taken
as the derivative of the corresponding local potential energy with respect to
strain. We assume that $\varepsilon_{\rm el}=0$ corresponds to the minimum of
the local basin and that $\varepsilon_{\rm el}=2\varepsilon^{*}$ corresponds
to the neighboring barrier. Consequently, the elastic stress vanishes at both
points and reaches a maximum $\sigma^{*}$ at
$\varepsilon_{\rm el}=\varepsilon^{*}$. This leads to the nonlinear elastic
constitutive relation
\begin{equation}
\sigma_{\rm el}(\varepsilon_{\rm el})
=
\sigma^{*}
\sin\left(
\frac{\pi}{2}
\frac{\varepsilon_{\rm el}}{\varepsilon^{*}}
\right)
=
E\frac{2\varepsilon^{*}}{\pi}
\sin\left(
\frac{\pi}{2}
\frac{\varepsilon_{\rm el}}{\varepsilon^{*}}
\right),
\label{elastic}
\end{equation}
where $\sigma^{*}
=
{2E\varepsilon^{*}}/{\pi}$.

Equation~\eqref{elastic} recovers linear elasticity for
$\varepsilon_{\rm el}\ll\varepsilon^{*}$, while the maximum at
$\varepsilon_{\rm el}=\varepsilon^{*}$ provides the elastic instability
associated with the stress overshoot.

For a prescribed stress below $\sigma^{*}$, Eq.~\eqref{elastic} has two
branches. Before the stress maximum,
\begin{equation}
\varepsilon_{\rm el}
=
\frac{2\varepsilon^{*}}{\pi}
\arcsin\left(
\frac{\sigma}{\sigma^{*}}
\right),
\label{elastic_ascending}
\end{equation}
whereas after the maximum,
\begin{equation}
\varepsilon_{\rm el}
=
\frac{2\varepsilon^{*}}{\pi}
\left[
\pi-
\arcsin\left(
\frac{\sigma}{\sigma^{*}}
\right)
\right].
\label{elastic_descending}
\end{equation}
The viscoplastic component of strain then follows directly from
\begin{equation}
\varepsilon_{\rm vp}(\varepsilon)
=
\varepsilon
-
\varepsilon_{\rm el}(\varepsilon).
\label{strain_decomposition}
\end{equation}

\subsection{Stress-relaxation and material time}

We introduce the relaxation dynamics associated with the viscous
element. For a conventional Maxwell element, stress-relaxation is governed by \cite{bergstrom2015mechanics}
\begin{equation}
\tau
\frac{d\varepsilon_{\rm el}}{dt}
+
\varepsilon_{\rm el}
=
0,
\end{equation}
which gives the Debye relaxation function
\begin{equation}
Y(t)
=
\exp\left(
-\frac{t}{\tau}
\right).
\label{eq:debye}
\end{equation}

Here, 
\begin{equation}
Y(t)=\sigma_{el}(t)/\sigma_{0}
\end{equation} 
and 

\begin{equation}
\sigma_{el}(t)=E \,\varepsilon_{\rm el}(t).
\end{equation}

Glassy materials, however, generally exhibit a broad spectrum of relaxation
times and are often described over an intermediate time regime by a
Kohlrausch-Williams-Watts (KWW) form \cite{kohlrausch1854theorie,williams1970non,PhysRevB_phillips,naumis2012bifurcation,liu2015mechanical_rev},
\begin{equation}
Y(t)
=
\exp\left[
-\left(
\frac{t}{\tau}
\right)^{\beta}
\right],
\label{eq:kww}
\end{equation}
where $\beta$ is the stretching exponent. The Debye limit is recovered for
$\beta=1$, whereas $\beta<1$ describes stretched relaxation.

It is useful to express these relaxation laws in terms of a dimensionless
material time $X$,
\begin{equation}
Y=\exp(-X).
\label{eq:material_relax}
\end{equation}
For the Debye case, $X=t/\tau$, while for the KWW form \cite{kohlrausch1854theorie,williams1970non},
$X=(t/\tau)^{\beta}$. More generally, $X$ can be regarded as a nonlinear but
monotonic clock describing the evolution of the relaxing degrees of freedom.
This construction follows directly from the relaxation introduced
in the model.

Since the KWW form is generally applicable only over an intermediate time
window, we introduce a material-time function that recovers Debye-like
behavior at both short and long times. We write
\begin{equation}
X_0(t)
=
\exp\left[
\beta_{\rm eff}(t)
\ln\left(
\frac{t}{\tau}
\right)
\right],
\label{eq:X0_time}
\end{equation}
where
\begin{equation}
\beta_{\rm eff}(t)
=
1-
(1-\beta_0)
\exp\left[
-\frac{
\left[
\ln(t/\tau)-P
\right]^2
}{Q}
\right].
\label{eq:beta_eff}
\end{equation}
Here, $\beta_0$ determines the maximum degree of stretching, $P$ determines
the position of the stretched-relaxation regime on a logarithmic time scale,
and $Q$ controls its width. Since
$\beta_{\rm eff}\simeq1$ at both very short and sufficiently
long times, the model approaches Debye-like dynamics in the two asymptotic
limits while retaining non-Debye relaxation over the intermediate regime.
The same phenomenological form is introduced in the underlying model to
connect experimentally accessible relaxation data to the nonlinear
stress-strain response. 

In Eq.~\eqref{eq:kww}, we use the subscript ``0" for $X_0$ to indicate that it is the material time for linear stress-relaxation. For deformation at constant strain rate,
$t=\varepsilon/\dot{\varepsilon}$, and hence we can express $X_0$ in terms of $\varepsilon$.
\begin{equation}
X_0(\varepsilon)
=
\exp\left[
\beta_{\rm eff}(\varepsilon)
\ln\left(
\frac{\varepsilon}
{\dot{\varepsilon}\tau}
\right)
\right],
\label{eq:X0_strain}
\end{equation}
with
\begin{equation}
\beta_{\rm eff}(\varepsilon)
=
1-
(1-\beta_0)
\exp\left[
-\frac{
\left[
\ln\left(
\varepsilon/(\dot{\varepsilon}\tau)
\right)-P
\right]^2
}{Q}
\right].
\end{equation}\label{beta_0}

\subsection{Nonlinear viscoplastic response}

We now generalize the dashpot response by expressing the viscoplastic stress in
terms of the material time, $X$,
\begin{equation}
\sigma_{\rm vp}
=
E
\frac{d\varepsilon_{\rm vp}}{dX}.
\label{eq:viscous_X}
\end{equation}
Because the elastic and viscoplastic elements are connected in series,
$\sigma_{\rm vp}=\sigma_{\rm el}$. Using Eq.~\eqref{elastic}, we therefore
obtain
\begin{equation}
E\frac{d\varepsilon_{\rm vp}}{dX}
=
E\frac{2\varepsilon^{*}}{\pi}
\sin\left[
\frac{\pi\varepsilon_{\rm el}}
{2\varepsilon^{*}}
\right].
\end{equation}
After canceling $E$, this can be written as
\begin{equation}
\frac{dX}{d\varepsilon_{\rm vp}}
=
\frac{\pi}
{
2\varepsilon^{*}
\sin\left[
\frac{\pi\varepsilon_{\rm el}}
{2\varepsilon^{*}}
\right]
}.
\end{equation}
Consequently, the nonlinear material time accumulated up to strain
$\varepsilon$ is
\begin{equation}
X(\varepsilon)
=
\int_0^{\varepsilon}
\frac{\pi}
{
2\varepsilon^{*}
\sin\left[
\frac{\pi\varepsilon_{\rm el}(\xi)}
{2\varepsilon^{*}}
\right]
}
\frac{d\varepsilon_{\rm vp}(\xi)}{d\xi}
\,d\xi.
\label{eq:X_nonlinear}
\end{equation}
This relation establishes a direct connection between the experimentally
observed nonlinear stress-strain response and the evolution of the viscous
material clock. 

To quantify the nonlinear modification of the viscous dynamics, we define the Eyring-type function \cite{eyring1936viscosity},
\begin{equation}
\phi(\varepsilon)
=
\frac{X(\varepsilon)}
{X_0(\varepsilon)}.
\label{eq:phi}
\end{equation}
Here, $X_0$ is the material time determined from the linear stress-relaxation
response, whereas $X$ is the material time required to reproduce the
nonlinear constant-rate deformation. Thus, $\phi$ measures the departure of
 nonlinear viscous dynamics from the equilibrium or linear-response
relaxation clock. In the linear elastic limit one requires
\begin{equation}
\phi(\varepsilon)\simeq1
\qquad
(\varepsilon\rightarrow0).
\end{equation}

At large strain, the elastic contribution approaches a finite value
$\varepsilon_Y$, while the viscoplastic strain carries the continuing deformation,
\[
\varepsilon_{\rm el}\simeq\varepsilon_Y,
\qquad
\varepsilon_{\rm vp}\simeq\varepsilon-\varepsilon_Y.
\]
At small strain,
$X_0\simeq\varepsilon/(\dot{\varepsilon}\tau)$ and
$\varepsilon_{\rm el}\simeq\varepsilon$. Expanding the sine function in
Eq.~\eqref{eq:X_nonlinear} then requires
\[
\frac{d\varepsilon_{\rm vp}}{d\varepsilon}
\simeq
\frac{\varepsilon}
{\dot{\varepsilon}\tau},
\]
and therefore
\begin{equation}
\varepsilon_{\rm vp}
\simeq
\frac{\varepsilon^2}
{2\dot{\varepsilon}\tau}
\qquad
(\varepsilon\rightarrow0).
\label{eq:visc_small}
\end{equation}
These small and large-strain asymptotes are the ones derived in the proposed
construction. 

The smooth strain components are calculated using the 
generalized Langmuir-T\'oth-type function \cite{toth1971state,terzyk2003developing},

\begin{equation}
\varepsilon_{\rm el}(\varepsilon)
=
\left[
\left(
\varepsilon
-
\frac{\varepsilon^2}
{2\dot{\varepsilon}\tau}
\right)^{-m}
+
\varepsilon_Y^{-m}
\right]^{-1/m},
\label{ansatz_el}
\end{equation}
together with,
\begin{equation}
    \varepsilon_{\rm vp}(\varepsilon)
=
\varepsilon
-
\varepsilon_{\rm el}(\varepsilon).
\label{ansatz_visc}
\end{equation}

The parameter $m$ controls the width of the crossover between the initial
elastic regime and the steady-flow regime; increasing $m$ produces a sharper
transition. We take $m>1$ so that the correction introduced by the
large-strain branch does not modify the required leading small-strain
asymptotics. The construction therefore satisfies
$\phi(\varepsilon)\simeq1$ in the elastic regime while allowing
$\phi(\varepsilon)$ to evolve nontrivially through yielding and approach a
finite value at large deformation. The same interpolation and limiting
conditions are used in the underlying model. 

The complete picture therefore separates three pieces of information:
the nonlinear elastic constitutive response, the reference relaxation clock
$X_0$ obtained from stress-relaxation measurements, and the nonlinear
material clock $X$ inferred from the constant-rate stress-strain response.
Their ratio $\phi=X/X_0$ provides a direct measure of the strain-induced
modification of viscoplastic relaxation and will be analyzed below using the
experimental data.

\section{Results and Discussion}
\begin{figure}[]
    \centering
    \includegraphics[width=0.88\linewidth,clip]{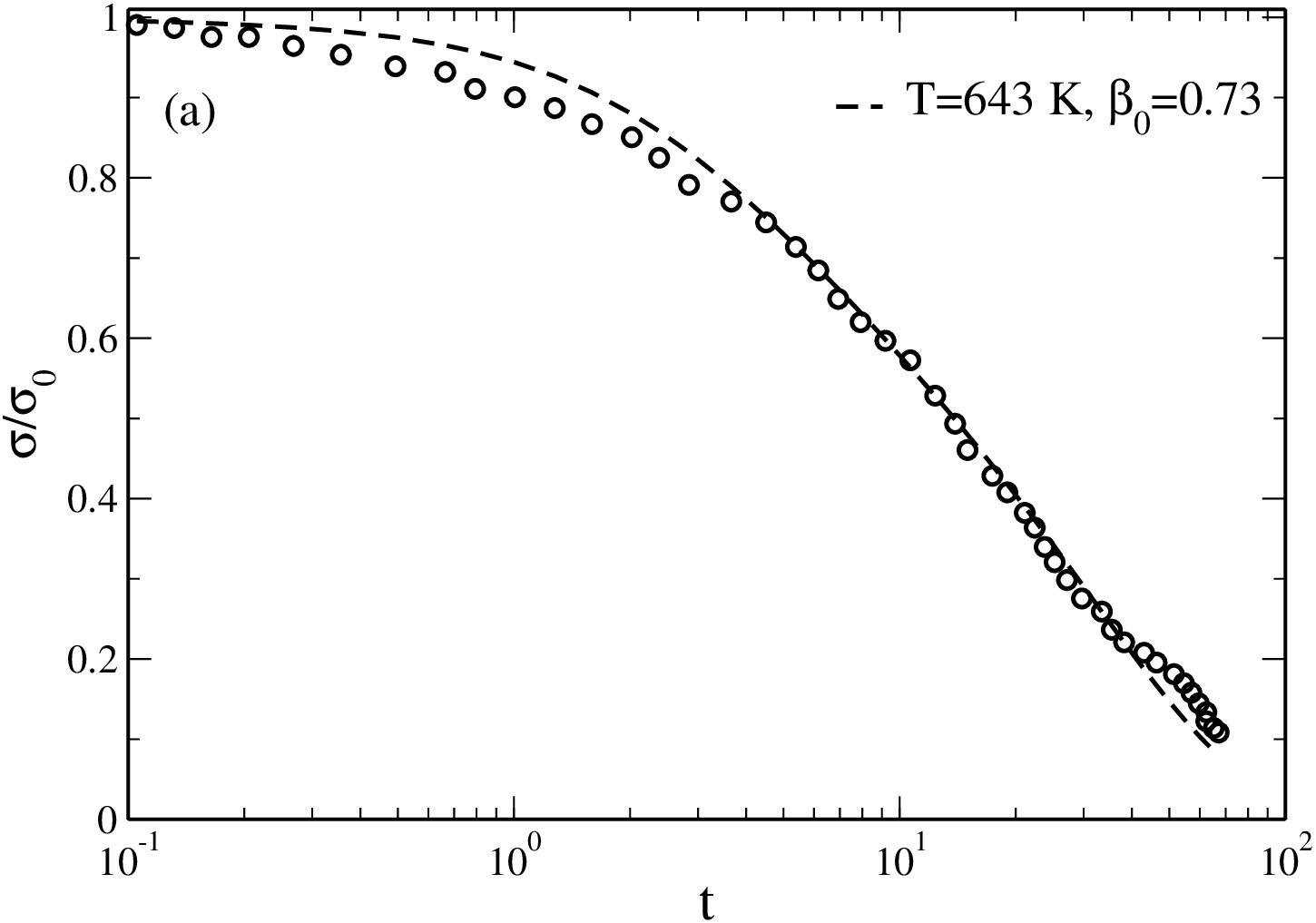}
    \includegraphics[width=0.88\linewidth,clip]{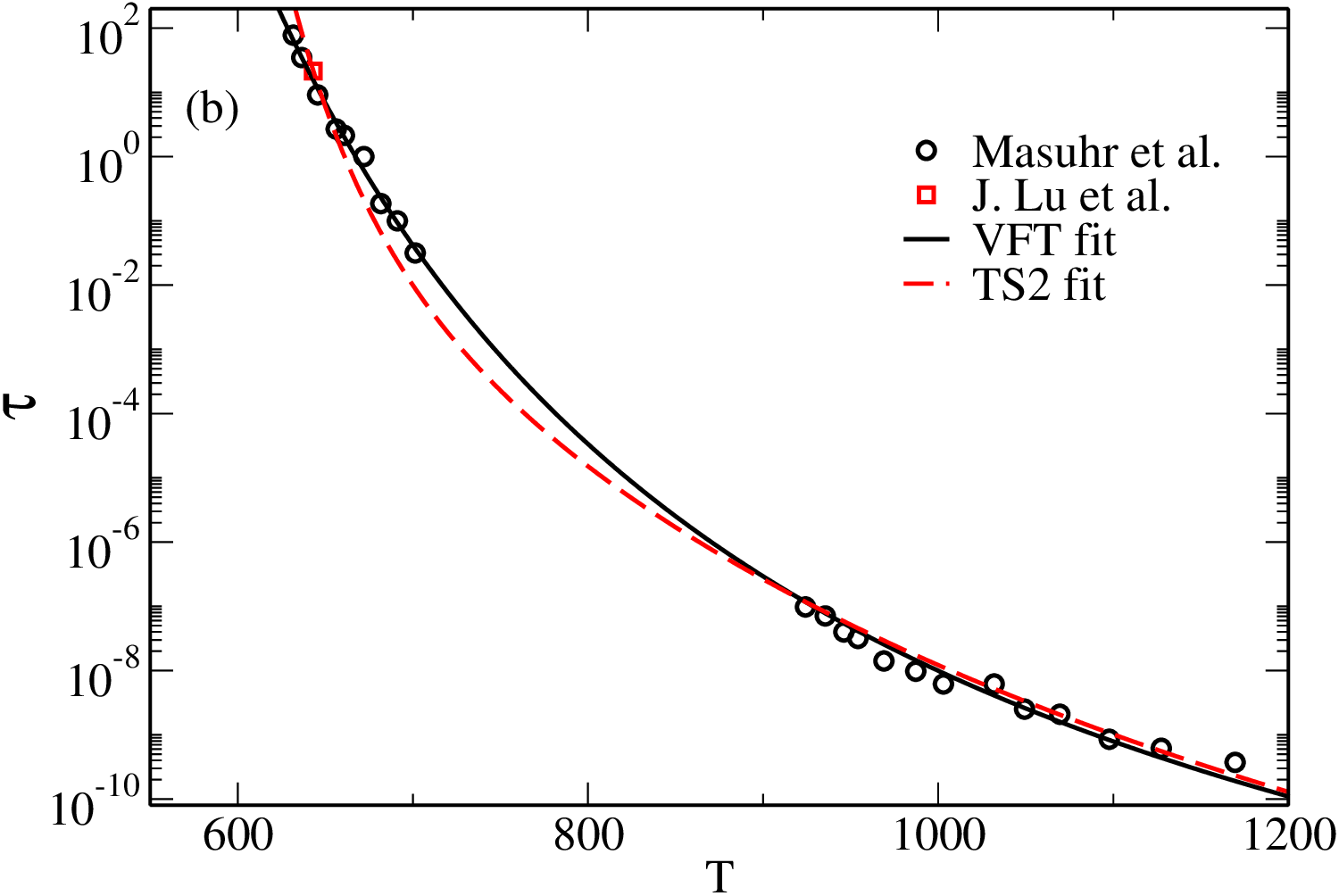}
    \caption{(a) Normalized stress as a function of time at temperature $T = 643~\mathrm{K}$. 
Symbols represent the experimental data, while the dashed lines correspond 
to stretched-exponential fits $\frac{\sigma}{\sigma_0}=e^{-X_0(t)}$ with a stretching exponent $\beta_{0} = 0.73$. 
(b) Relaxation time as a function of temperature. Symbols represent the 
experimental values \cite{masuhr1999thermodynamics,masuhr1999time,lu2003deformation}, while the solid and dashed lines correspond to the 
VFT and TS2 fits, respectively.
}\label{stress_time}
\end{figure}

We have used a typical metallic glass $\mathrm{Zr}_{41.2}\mathrm{Ti}_{13.8}\mathrm{Cu}_{12.5}\mathrm{Ni}_{10}\mathrm{Be}_{22.5}$ (Vitreloy 1), with a glass transition temperature of $T_g = 623 \pm 2~\mathrm{K}$ \cite{lu2003deformation,glass_transition_1993}, as a model material for our theoretical model and calculations.
We first determine the relaxation time entering the material-time description from the independent stress-relaxation data shown in Fig.~\ref{stress_time}. Figure~\ref{stress_time}(a) shows the normalized stress-relaxation, $\sigma/\sigma_0$, of the metallic glass at $T=643~\mathrm{K}$ \cite{lu2003deformation}. The experimental data show a clear non-Debye decay and are well described by a stretched-exponential relaxation with $\beta_{0}=0.73$ \cite{PhysRevB_phillips,liu2015mechanical_rev}. The value $\beta_0<1$ indicates that the relaxation cannot be described by a single Maxwell relaxation time, but instead involves a broad distribution of structural relaxation processes characteristic of the glassy state. Figure~\ref{stress_time}(b) shows the corresponding relaxation time as a function of temperature. The relaxation time changes over several orders of magnitude with temperature and is well described by both the Vogel-Fulcher-Tammann (VFT) and TS2 fits. From these data, we obtain relaxation times of $\tau = 22~\mathrm{s}$ at $T = 643~\mathrm{K}$ and $\tau = 2.7~\mathrm{s}$ at $T = 663~\mathrm{K}$. These relaxation times are relevant to the corresponding stress--strain measurements and are used to separate the elastic and viscoelastic contributions to the deformation. This strong temperature dependence is important because the competition between the imposed deformation time scale and the intrinsic relaxation time determines how the total strain is divided between elastic and viscoplastic contributions.

The temperature dependence of the structural relaxation time \(\tau\) has been analyzed using both the VFT equation and the TS2 model. The VFT equation is given by \cite{RevModPhys_dyre_2006},

\begin{equation}
\log_{10}\tau(T)
=
\log_{10}\tau_0+
\frac{B}{(T-T_0)},
\end{equation}
where \(\tau_0\) is the microscopic relaxation time, \(B\) is a constant, and \(T_0\) is the VFT temperature. Fitting the relaxation-time data gives \(\log_{10}(\tau_0)=-16.789\), corresponding to \(\tau_0=1.627\times10^{-17}\,\mathrm{s}\), \(B=6132.92\,\mathrm{K}\), and \(T_0=301.97\,\mathrm{K}\), with a root-mean-square error (RMSE) of 0.20 in \(\log_{10}(\tau)\).

We have also used the TS2 model that describes the temperature dependence of the relaxation time through a crossover between two activation regimes \cite{TS2_model_2026},

\begin{equation}
\log_{10}\left(\frac{\tau(T)}{\tau_{\mathrm{el}}}\right)
=
\frac{\bar{E}_1}{\bar{T}}
+
\frac{\bar{E}_2-\bar{E}_1}{\bar{T}}
\frac{1}{1+\exp\left[\Delta\bar{S}\left(1-\frac{1}{\bar{T}}\right)\right]},
\end{equation}
where $\bar{T}=T/T_X$, $T_X$ is the crossover temperature, $\bar{E}_1$ and $\bar{E}_2$ are dimensionless activation energies, and $\Delta\bar{S}$ governs the crossover. The microscopic relaxation time $\tau_{\mathrm{el}}$ is related to the Arrhenius relaxation time $\tau_A$ through $\log_{10}(\tau_{\mathrm{el}})=\log_{10}(\tau_A)-17.7$. Here, $\bar{E}_1=50$, $\bar{E}_2=150$, and $\Delta\bar{S}=16$ are universal values \cite{TS2_model_2026}. The TS2 model therefore has two adjustable parameters, $T_X$ and $\tau_A$. For metallic glass, the fit gives $T_X=542\,\mathrm{K}$ and $\log_{10}(\tau_A)=-2.0$, with a RMSE of $0.24$ in $\log_{10}(\tau)$.

\begin{figure}
 \includegraphics[width=0.88\linewidth,clip]{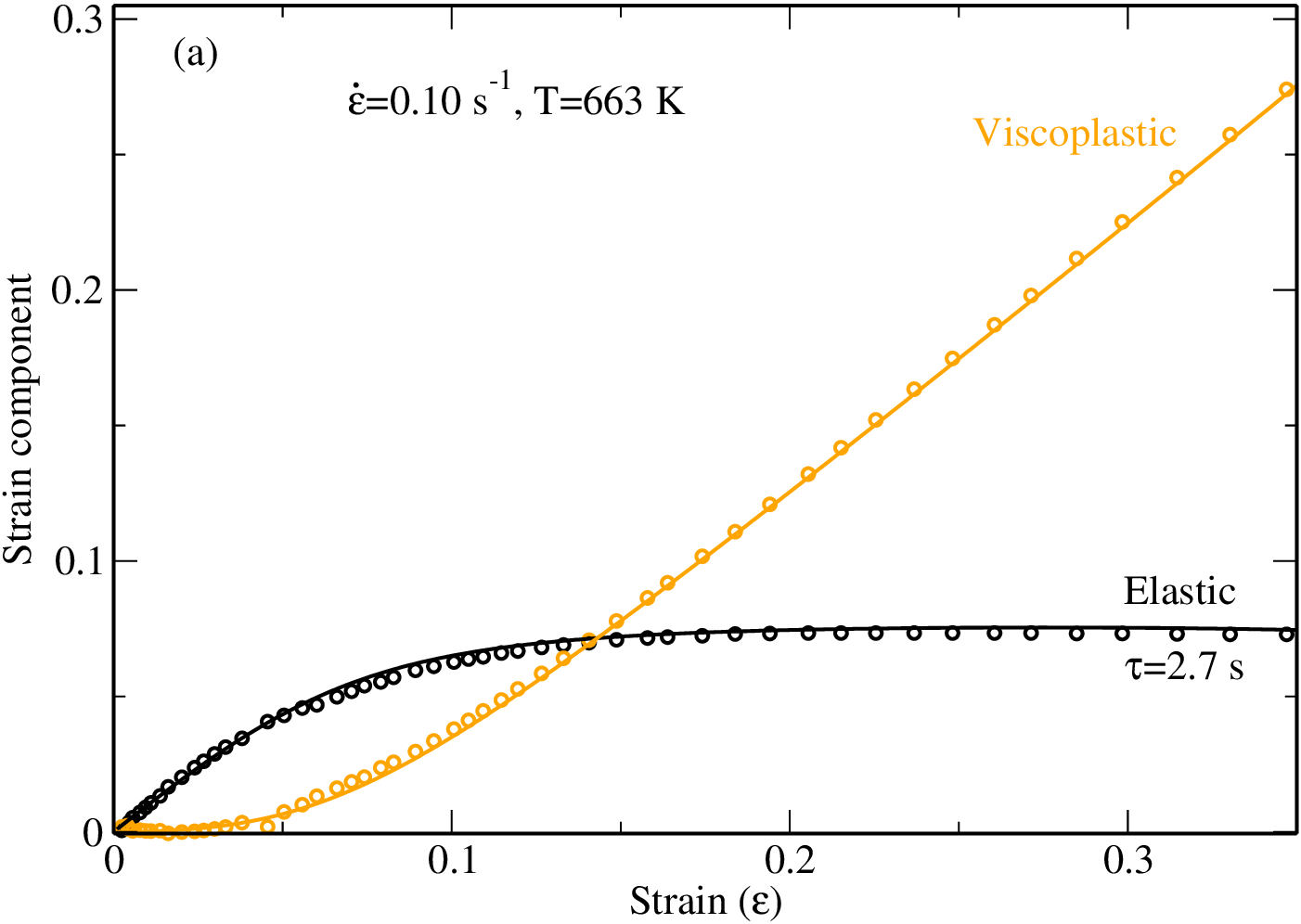}
    \includegraphics[width=0.88\linewidth,clip]{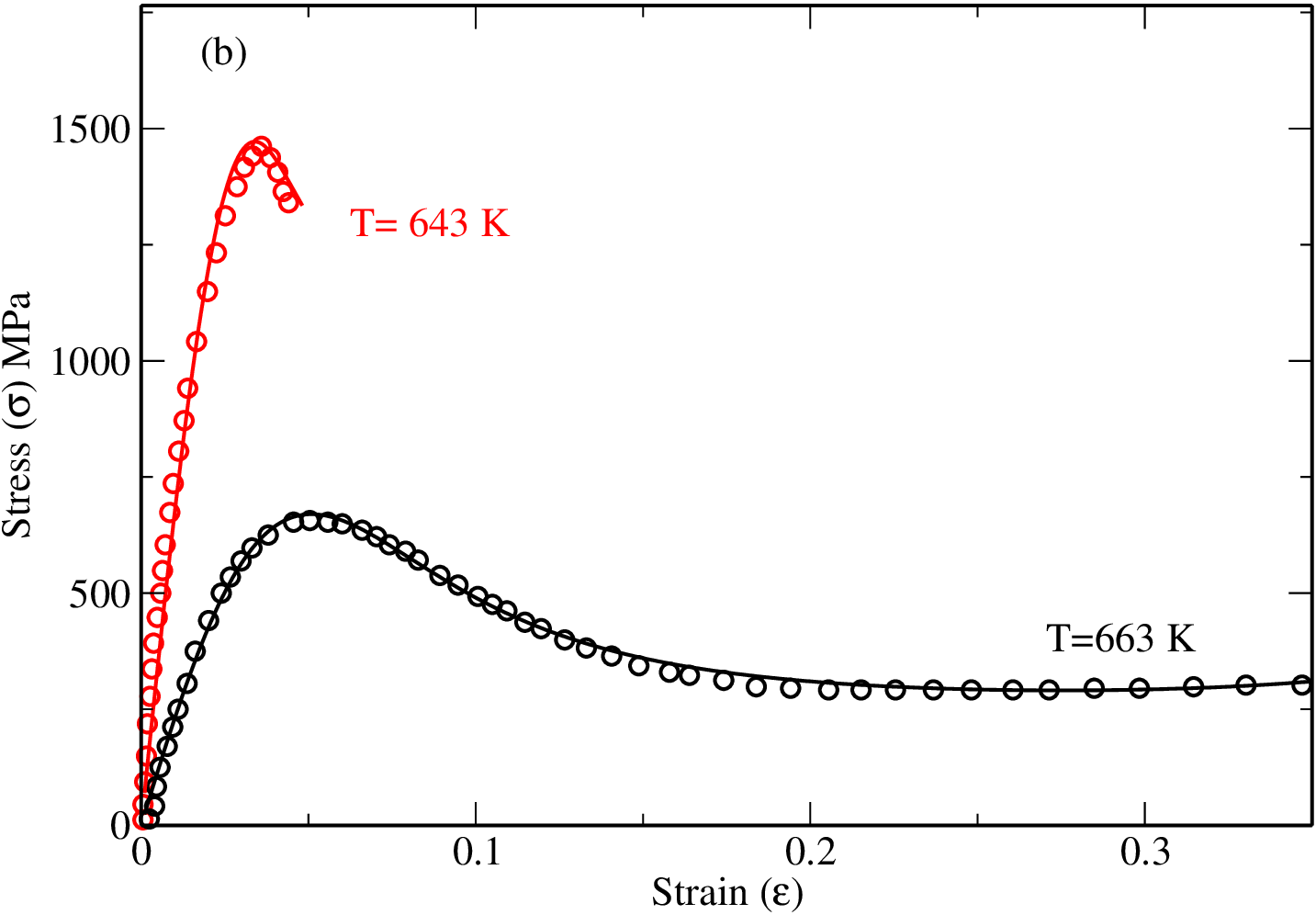}
    \caption{(a) Elastic and viscoplastic strain components as functions of the total strain at $T=663~\mathrm{K}$ and $\dot{\varepsilon}=0.10~\mathrm{s}^{-1}$. Open circles represent the elastic and viscoplastic strain components obtained from the experimental data using $\sigma/\sigma^*$ and $\epsilon^*$ in Eqs.~\eqref{elastic_ascending}-\eqref{strain_decomposition}, while solid lines represent the corresponding strain components calculated using the GL-based strain decomposition using Eqs.~\eqref{ansatz_el}-\eqref{ansatz_visc}, with $\varepsilon_Y$, $m$, and  $\tau$    listed in Table~\ref{table}. (b) Stress-strain curves of the metallic glass at $T=643~\mathrm{K}$ and $663~\mathrm{K}$ for $\dot{\varepsilon}=0.10~\mathrm{s}^{-1}$. Open circles represent the experimental data, while the solid lines are calculated from the nonlinear stress model, Eq.~\ref{elastic}  using the corresponding elastic strain component.}
 \label{Stress_temp}
\end{figure}

Figure~\ref{Stress_temp}(a) shows the elastic and viscoplastic strain components at temperature
$T=663~\mathrm{K}$ and $\dot{\varepsilon}=0.10~\mathrm{s^{-1}}$.
The symbols are obtained directly from the experimental stress-strain curve
using Eqs.~\eqref{elastic_ascending}-\eqref{strain_decomposition}, whereas
the solid lines are calculated using the generalized Langmuir-T\'oth-type
saturation function given by Eq.~\eqref{ansatz_el}. At small deformation, the elastic strain follows the total strain almost linearly, while the viscoplastic contribution remains very small. This is consistent with the model, where the initial deformation is mainly elastic. With increasing strain, the viscoplastic contribution gradually increases, while the elastic strain bends over and approaches a limiting value. This indicates the crossover from a mainly elastic response to a regime where an increasing part of the imposed deformation is accommodated by viscous flow. At large strain, the elastic contribution becomes nearly constant, whereas the viscoplastic strain continues to increase with the total strain. The crossover between these two regimes is controlled by the relaxation time $\tau$ and the  parameter $m$, which determines how sharply the elastic strain approaches its maximum value and listed in Table \ref{table}.

The corresponding stress-strain response is shown in Fig.~\ref{Stress_temp}(b) for the
metallic glass at $T=643~\mathrm{K}$ and $663~\mathrm{K}$ with
$\dot{\varepsilon}=0.10~\mathrm{s^{-1}}$. Open circles represent the
experimental data, while the solid lines are calculated from the nonlinear
stress relation in Eq.~\eqref{elastic} using the corresponding elastic strain
component obtained from Eq.~\eqref{ansatz_el}. At small strain, $\varepsilon_{\mathrm{el}}\simeq\varepsilon$, and the response follows the initial elastic loading branch. As the deformation increases, the elastic strain approaches the characteristic value $\varepsilon^{*}$, where the stress reaches its maximum value $\sigma^{*}$ and the stress overshoot appears. Beyond this point, the elastic strain enters the descending branch of the nonlinear elastic response, while an increasing fraction of the imposed deformation is transferred to the viscoplastic component. This produces the decrease of the stress after the overshoot. At  larger strain, the elastic strain approaches its limiting value $\varepsilon_Y$, and the stress approaches the steady yielding regime, while the continuing deformation is mainly accommodated by the viscoplastic component.

The comparison between $T=643~\mathrm{K}$ and $663~\mathrm{K}$ in
Fig.~\ref{Stress_temp}(b) clearly shows the effect of temperature on the relaxation and
nonlinear elastic response. At $T=643~\mathrm{K}$, the relaxation time is
$\tau\simeq22~\mathrm{s}$, whereas at $663~\mathrm{K}$ it decreases to
approximately $2.7~\mathrm{s}$, indicating much faster structural relaxation
during deformation. At the same time, the characteristic stress
$\sigma^{*}$ decreases from $1472$ to $670~\mathrm{MPa}$, which leads to a
strong reduction of the stress overshoot. The yield strain changes from
$\varepsilon_Y=0.045$ to $0.079$, while the crossover parameter $m$ decreases
from $6$ to $3.3$, corresponding to a smoother elastic-viscoplastic crossover at
the higher temperature. Thus, the reduced stress overshoot at $663~\mathrm{K}$
results from the combined effect of faster structural relaxation and the
temperature dependence of the nonlinear elastic response.

Results show that the temperature dependence of the nonlinear stress-strain response can be directly connected to the relaxation time obtained independently from the stress-relaxation data. The relaxation gives the characteristic material time, while the nonlinear elastic relation determines the stress associated with the instantaneous elastic part of the deformation. Their combination through the elastic-viscoplastic strain decomposition provides a consistent description of the initial elastic response, the stress overshoot, the post-yield stress decrease, and the steady viscoplastic flow of the metallic glass.

\begin{table}[]
\caption{Model parameters used to describe the stress-strain response of the metallic glass. Here, $\sigma^{*}$ is the characteristic stress scale, $\varepsilon_Y$ is the yield strain, $\tau$ is the relaxation time, $m$ is the stress-dependent activation parameter, $\beta_0$ is the stretching exponent, and $\dot{\varepsilon}$ is the applied strain rate. The first two rows correspond to temperature $T=643~\mathrm{K}$ at two strain rates, while the last row corresponds to $T=663~\mathrm{K}$ at $\dot{\varepsilon}=0.10~\mathrm{s^{-1}}$.}
\vspace{0.5cm}
\centering
\resizebox{0.99\columnwidth}{!}{%
\begin{tabular}{lcccccccc}
\hline
Material &$\dot{\varepsilon}$ (s$^{-1}$) & $\sigma^{*}$$\mathrm{(MPa)}$ & $\varepsilon_{Y}$  & $\tau$$ (s$) & $m$ & $\beta_0$ &  \\
\hline
$T=643~\mathrm{K}$ & $0.10$ & $1472$ & $0.045$ & $22$ & $6$ &  $0.73$ & \\
           &$0.032$    & $1143$ & $0.076$ & $22$ & $4$ & $0.73$ & \\

               \\
\hline

$T=663~\mathrm{K}$     &$0.10$          & $670$ & $0.079$ & $2.7$ & $3.3$ & $0.75$ & \\
\hline\end{tabular}}
\label{table}
\end{table}

\begin{figure}
    \includegraphics[width=0.88\linewidth,clip]{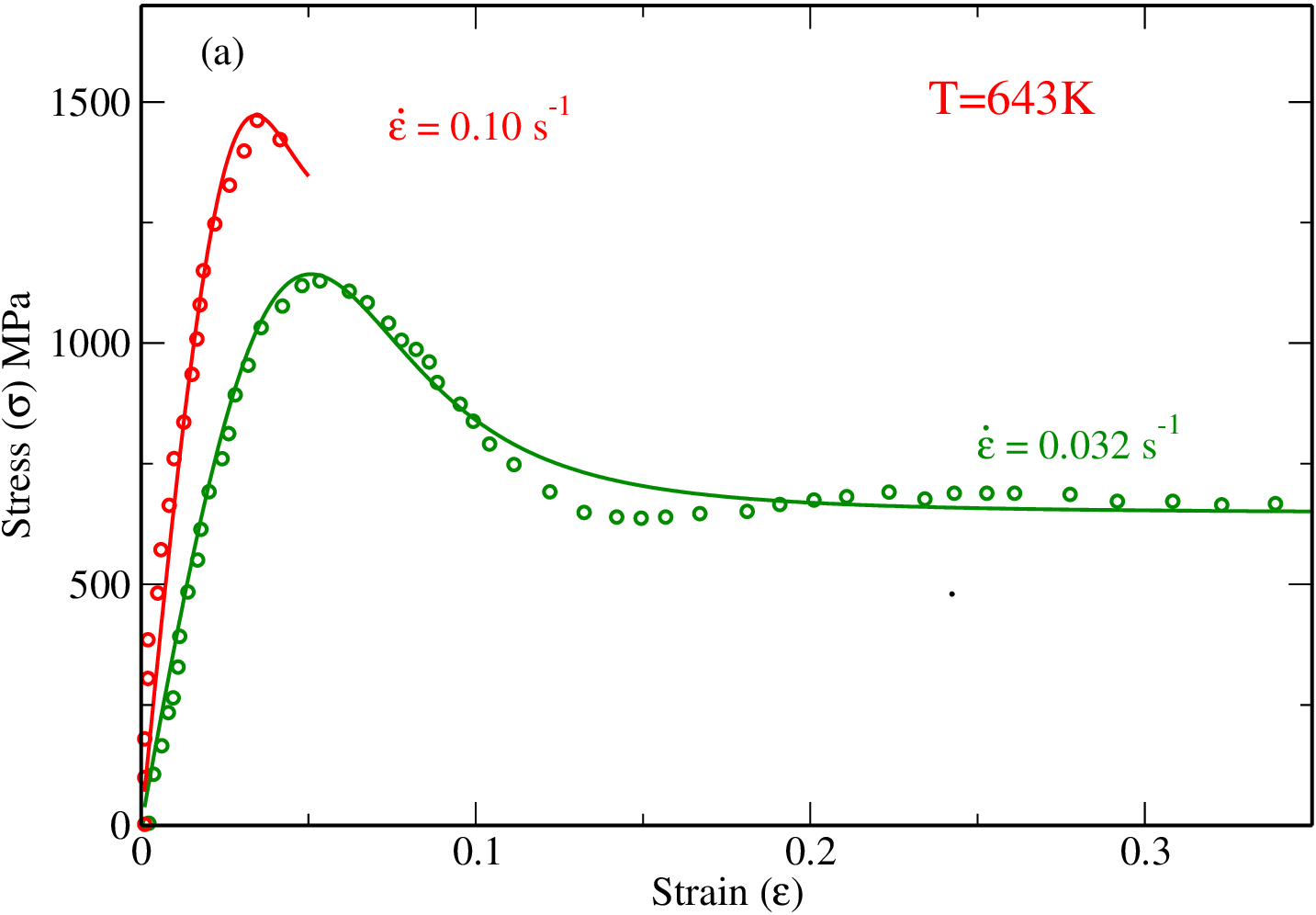}
     \includegraphics[width=0.88\linewidth,clip]{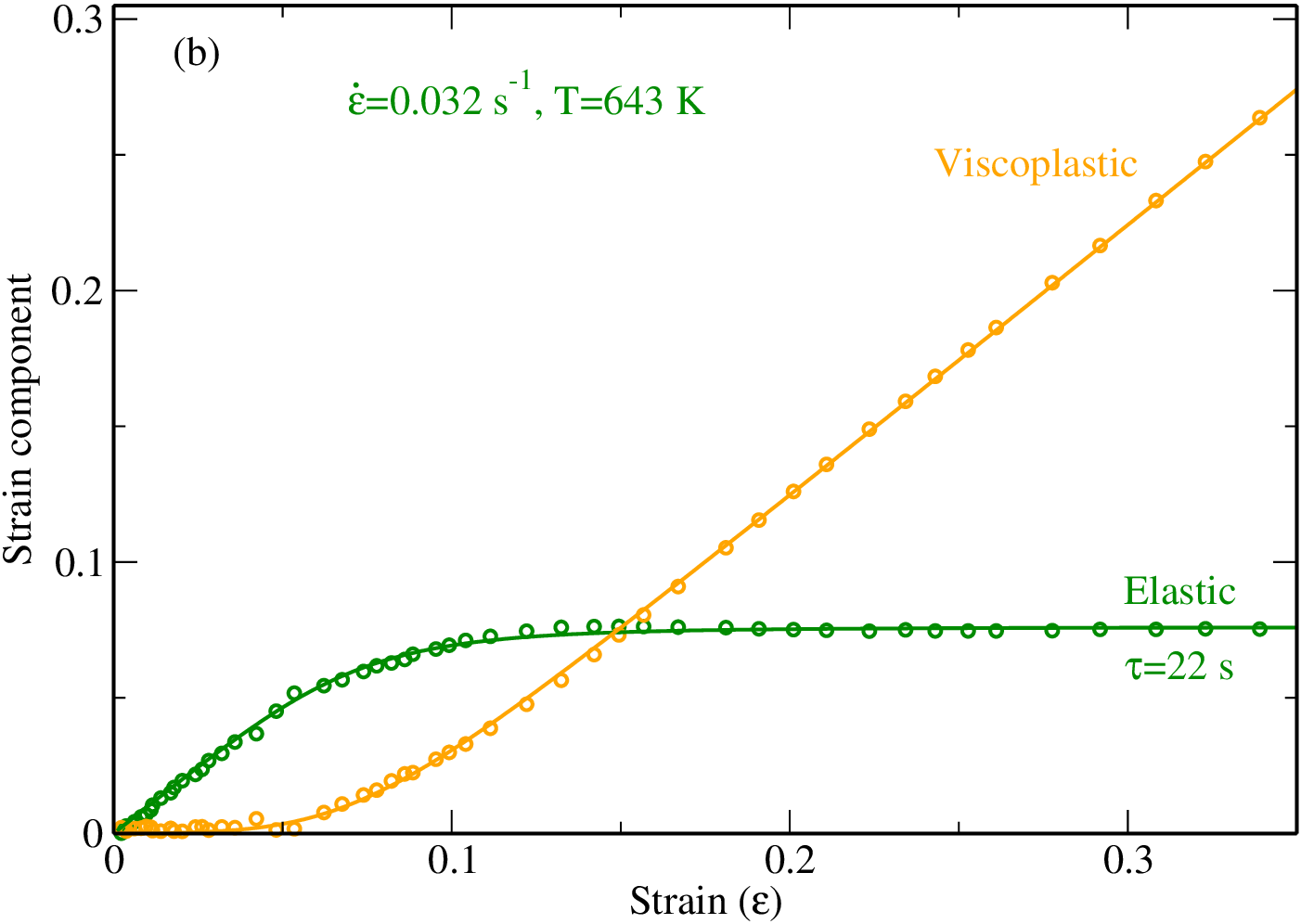}
       
    \caption{(a) Stress-strain curves of a metallic glass at $T=643~\mathrm{K}$ for strain rates $\dot{\varepsilon}=0.10$ and $0.032~\mathrm{s}^{-1}$. Open circles represent the experimental data, while the solid lines are calculated using the nonlinear stress model. (b) Elastic and viscoplastic strain components as functions of the total strain at $T=643~\mathrm{K}$ for $\dot{\varepsilon}=0.032$.}
 \label{Stress_srainrate}
\end{figure}

Figure~\ref{Stress_srainrate} further examines the strain-rate dependence of the nonlinear response at a fixed temperature, $T=643~\mathrm{K}$. Figure~\ref{Stress_srainrate}(a) shows the stress-strain curves for $\dot{\varepsilon}=0.10$ and  $0.032 ~\mathrm{s^{-1}}$. A clear stress overshoot is observed for these two strain rates, followed by a decrease toward the steady-flow regime. The magnitude of the stress overshoot strongly depends on the applied strain rate. At the highest strain rate, $\dot{\varepsilon}=0.10~\mathrm{s^{-1}}$, the material develops the largest maximum stress because the imposed deformation is fast compared with structural relaxation. As the strain rate is decreased, the material has more time to relax during deformation, and consequently both the maximum stress and the subsequent flow stress decrease. The position of the stress maximum also shifts toward larger strain with decreasing strain rate. The solid lines obtained from the nonlinear stress model reproduce the experimental data over the complete deformation range, including the initial elastic regime, the stress overshoot, and the post-yield flow response. Figures~\ref{Stress_temp} and \ref{Stress_srainrate} show a clear brittle-to-ductile transition in between temperature $T = 643\,\mathrm{K}$ and $T = 663\,\mathrm{K}$ at a strain rate of $0.10\,\mathrm{s}^{-1}$, characterized by a shift from abrupt failure to sustained plastic deformation~\cite{soft_matter_2026}.

The corresponding elastic-viscoplastic strain decomposition is shown in Figs.~\ref{Stress_srainrate}(b) for $\dot{\varepsilon}=0.032\mathrm{s^{-1}}$. The deformation is mainly elastic at small strain, while the viscoplastic contribution is initially very small. With increasing strain, the elastic component gradually approaches a limiting value, whereas the viscoplastic component continuously increases and eventually becomes the dominant contribution to the total deformation. Therefore, after yielding, most of the additional applied strain is accommodated through viscoplastic deformation, while the elastic strain remains nearly saturated.

Overall, Fig.~\ref{Stress_srainrate} shows that the strain-rate dependence of the stress overshoot can be understood through the competition between elastic loading and viscoplastic relaxation. At high strain rate, deformation is accumulated faster than the material can relax, resulting in stronger elastic stress storage and a larger overshoot. At lower strain rate, structural relaxation becomes more effective during loading, reducing the elastic stress and allowing a larger fraction of the deformation to be accommodated viscously. Together with Figs.~\ref{Stress_temp} and \ref{Stress_srainrate}, these results show that the same elastic-viscoplastic model can describe both the temperature and strain-rate dependence of the nonlinear response of the metallic glass.

\begin{figure}
    \includegraphics[width=0.95\linewidth,clip]{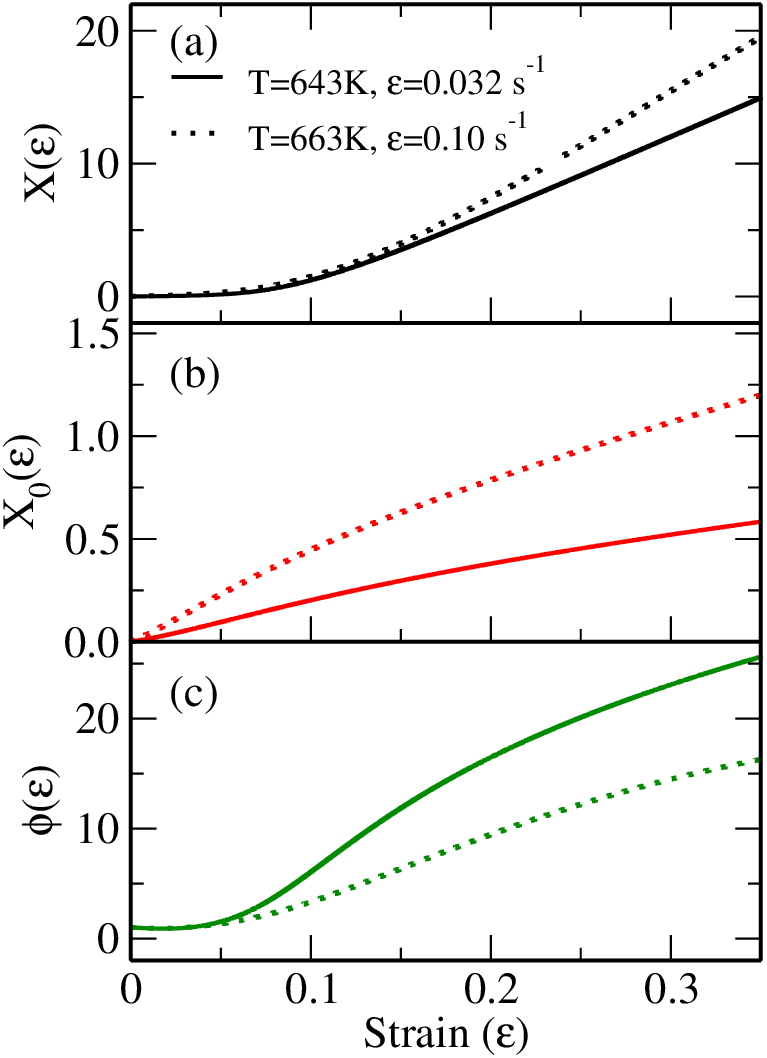}
  \caption{(a) (a) Nonlinear material-time variable $X(\varepsilon)$ obtained from the elastic viscoplastic strain decomposition using Eq.~\eqref{eq:X_nonlinear}. (b) Reference material-time function $X_0(\varepsilon)$ calculated from the stress-relaxation response using Eq.~\eqref{eq:X0_strain}. (c) Eyring function $\phi(\varepsilon)=X(\varepsilon)/X_0(\varepsilon)$ as a function of strain. In the small-strain regime, $\phi(\varepsilon)\simeq 1$, indicating that the material-time evolution remains close to its reference behavior. Around the onset of yielding, $\phi(\varepsilon)$ starts to increase and continues to grow with strain, reflecting the increasing deviation from the small-strain relaxation dynamics and the enhanced role of viscoplastic deformation. Solid lines represent the calculations at $T=643~\mathrm{K}$ and $\dot{\varepsilon}=0.032~\mathrm{s}^{-1}$, while dotted lines represent those at $T=663~\mathrm{K}$ and $\dot{\varepsilon}=0.10~\mathrm{s}^{-1}$.}
 \label{Eyring}
\end{figure}
Figure~\ref{Eyring} shows the evolution of the material-time quantities under the same deformation conditions as in Figs.~\ref{Stress_temp} and \ref{Stress_srainrate}, at temperatures $T = 663~\mathrm{K}$ and $643~\mathrm{K}$, with strain rates $\dot{\varepsilon} = 0.10~\mathrm{s}^{-1}$ and $0.032~\mathrm{s}^{-1}$, respectively. The reference material-time function $X_0(\varepsilon)$ is obtained from the stress-relaxation response using $\tau=22~\mathrm{s}$, $\beta_0=0.73$, $P=-0.73$, and $Q=5$. The dotted lines correspond to $T=663~\mathrm{K}$ and $\dot{\varepsilon}=0.10~\mathrm{s}^{-1}$, for which $X_0(\varepsilon)$ is obtained using $\tau=2.7~\mathrm{s}$, $\beta_0=0.75$, $P=0.0$, and $Q=5$ \cite{PhysRevB_phillips,liu2015mechanical_rev}. In both cases, $X_0(\varepsilon)$ increases monotonically with strain, representing the reference relaxation clock derived from the stress-relaxation response.

The nonlinear material-time variable $X(\varepsilon)$ is calculated independently from the elastic-viscoplastic strain decomposition. At small strain, $X(\varepsilon)$ remains close to the reference material time, consistent with the mainly elastic response in this regime. With increasing strain, particularly close to the onset of yielding, $X(\varepsilon)$ starts to increase much faster than $X_0(\varepsilon)$. This reflects the increasing contribution of the viscoplastic strain after the elastic component begins to saturate. Therefore, the nonlinear material time progressively deviates from the relaxation behavior obtained from the small-deformation stress-relaxation response.

This deviation is more clearly seen from the Eyring-type function,
$\phi(\varepsilon)=X(\varepsilon)/X_0(\varepsilon)$, shown in Fig.~\ref{Eyring}(c). In the small-strain regime, $\phi(\varepsilon)\simeq1$, indicating that the nonlinear deformation initially follows the reference material-time dynamics. Around the strain range where yielding starts, $\phi(\varepsilon)$ begins to increase and then continues to grow with strain. This increase indicates that the material-time evolution during nonlinear deformation becomes progressively different from that obtained from linear stress-relaxation. This behavior is directly connected to the transfer of deformation from the elastic component to the viscoplastic component during and after yielding.

Thus, Fig.~\ref{Eyring} provides a direct connection between the stress-relaxation dynamics and the nonlinear stress-strain response. The function $X_0(\varepsilon)$ represents the reference material clock obtained from relaxation measurements, whereas $X(\varepsilon)$ represents the material clock required to describe the nonlinear deformation. Their ratio $\phi(\varepsilon)$ therefore gives a direct measure of the nonlinear modification of the relaxation dynamics. Its value close to unity at small strain and its increase around yielding are consistent with the crossover from predominantly elastic deformation to a regime increasingly dominated by viscoplastic flow.

To estimate the activation/correlation volume \(V^*\), we analyze the strain dependence of the Eyring function \(\phi\) at temperatures of 643 K and 663 K. For these two temperature, \(\ln\phi\) is fitted to a quadratic form, \(\ln\phi=k\epsilon^2\), over the strain interval \(0<\epsilon\leq0.10\). In this regime, anharmonic contributions to the potential energy and the difference between total and elastic strain are assumed to be negligible to a first approximation. The effective modulus is estimated from the characteristic stress and strain as \(E=\pi\sigma^*/(2\epsilon^*)\), modulus \(E=38.20\,\mathrm{GPa}\) at 643 K and \(E=23.92\,\mathrm{GPa}\) at 663 K. Assuming that the fitted quadratic coefficient corresponds to the elastic-energy contribution \(\Delta U=EV^*\epsilon^2/2\), the effective volume is calculated as \(V^*=2k\, k_{\mathrm B}T/E\)~\cite{jiang_2015origin,fan2014thermally,puthoff2007change}, where \(k_{\mathrm B}=1.380649\times10^{-23}\,\mathrm{J\,K^{-1}}\) is Boltzmann's constant. We found that the  activation volumes are \(V^*=90.31\,\mathrm{\AA}^3\) at 643 K and \(V^*=97.37\,\mathrm{\AA}^3\) at 663 K.

\section{Summary and Conclusions}

We developed a nonlinear visco-elasto-plastic model for the stress-strain response of metallic glasses by connecting nonlinear deformation with the material relaxation time. The total strain is decomposed into elastic and viscoplastic contributions within a Maxwell-type description. The nonlinear elastic response gives the stress overshoot and yielding, while the viscoplastic contribution is connected to the material time obtained independently from stress-relaxation data.

Our framework describes the crossover from the initial elastic response to viscoplastic flow and captures the observed dependence on temperature and strain rate. A longer relaxation time leads to stronger elastic stress storage and a more pronounced stress overshoot, whereas faster relaxation increases the viscoplastic contribution and reduces the overshoot. The reference material-time functions $X_0(\varepsilon)$ and nonlinear material-time variable $X(\varepsilon)$ connect the linear relaxation dynamics with the nonlinear deformation. In particular, Eyring-type function $\phi(\varepsilon)$ remains close to unity at small strain, indicating that the nonlinear deformation initially follows the reference material-time dynamics and increases around yielding, showing the change in relaxation dynamics during nonlinear deformation. The estimated activation volumes provide insight into the thermally activated deformation mechanisms. Together with material-time analysis, these results highlight the interplay between molecular relaxation, elastic energy storage, and viscoplastic flow, providing a consistent description of the nonlinear mechanical response across different temperatures and strain rates.

The present model successfully describes the mechanical response of a metallic (zirconium-based) metallic glass. In the future, we plan to extend the model to other amorphous materials, specifically polymer glasses such as polycarbonate (PC),  poly(methyl methacrylate) (PMMA), and epoxy thermosets~\cite{siviour2016high,choi2026solid}. This will require incorporating the effects of polymer strain hardening and will be the subject of future work.

\section*{acknowledgments}
AS and AZ are grateful for the funding from the European Union through Horizon Europe ERC Grant number: 101043968, “Multimech”.

\bibliography{manuscript}

\end{document}